\documentstyle[aps]{revtex}
\begin{document}

\title{Charged Black Holes In Quadratic Theories}

\author {A. Economou and C.O. Lousto}

\address{
  Fakult\"at f\"ur Physik der
  Universit\"at Konstanz,
  Postfach 5560,
  D - 78434 Konstanz, Germany
}

\maketitle

\begin{abstract}
 We point out that in general the
Reissner-Nordstr\"om (RN) charged black holes of general relativity
are not solutions of the four dimensional quadratic
gravitational theories.
They are, e.g., exact solutions of the $R+R^2$
quadratic theory but not of a theory where a $R_{ab}R^{ab}$ term is
present in the gravitational Lagrangian.
In the case where such a non linear curvature
term  is present with sufficiently small coupling, we obtain an
approximate solution for a charged black hole of charge $Q$ and mass $M$.
For $Q\ll M$ the validity of this solution extends down to the horizon.
This allows us to explore the thermodynamic properties of the quadratic
charged black hole and we find that, to our approximation,
its thermodynamics is identical to that of a RN black hole.
However our black hole's entropy is not equal to the one fourth of
the horizon area.
Finally we extend our analysis to the rotating charged black hole
and qualitatively similar results are obtained.
\end{abstract}

\pacs{PACS numbers: 04.60.+n}

\section{Introduction}
In the past few years there has been a considerable interest for black
hole (BH) solutions and their properties.
One of the reasons is the developments in our modern physical theories,
in particular in gauge field theories and in string theory.
On the one hand these developments  have enriched
the palette of known fields that probably play some role in Nature
and of course couple to gravity
because of the universal character of the gravitational interaction.
On the other hand, it is expected that Einstein's
theory of gravity  gets effectively modified, at some higher energy
scale.
We may have in the gravitational Lagrangian non linear curvature
terms and the effective coupling with matter need not to be minimal.
Also the spacetime dimensionality may be higher than four
with the extra dimensions compactified by some appropriate mechanism.

Incorporation of such elements in gravitation turns out to be
particularly interesting and has already given rise to new and,
sometimes, surprising consequences for  black hole physics.
In particular the discovery of a new family of BH solutions for
Yang-Mills fields \cite{Bartnik88} has set us free from the
so-called ``no-hair'' conjecture and its related theorems.
We now know that black holes do not necessarily belong to the
standard black hole families of solutions, namely the
Schwarzschild, Reissner-Nordstr\"om (RN) or the Kerr-Newman (KN) ones,
and furthermore that in general they are not unique.
In other words, they may have ``hair''.
The properties of the Yang-Mills fields that are responsible for the
existence of new black hole solutions have been extensively analyzed in
Ref.\ \cite{Sudarsky92}.
For comments on the way classical no-hair theorems
can be avoided by the presence of some fields,
e.g. the dilaton in effective theories inspired on string theory,
see Ref.\ \cite{Alwis92}.

With the gravity theories modified one may of course expect that
also the thermodynamical properties of black hole laws
for generalized black holes, may significantly
differ from that of standard black holes. However several issues do
survive. E.g. the first law is
still valid for Einstein-Yang-Mills black hole solutions \cite{Sudarsky92}.
In this paper we are interested in a somehow accidental property of standard
black holes, sometimes referred in the literature as the area law.
It is the simple proportionality between the black hole entropy $S$
and its geometric intrinsic quantity, the horizon's area $A$,
namely the famous $S=A/4$ relation (in appropriate units).
Interestingly enough,
this relationship is still valid for some generalized black holes,
but there are known cases where this is not true. The underlying pattern, if
any, is not so clear.
For references on  cases where this relationship
is not valid see \cite{Visser93}. In the same reference an interesting
entropy formula is derived which, in some sense, systematizes the known
results.

Of particular interest for this paper are cases where the
gravitational Lagrangian  contains non linear curvature terms.
For higher dimensionality $d>4$ it seems that the area law is
generically not valid.
It is interesting that for Lovelock gravity, the entropy is still given by
an intrinsic geometric quantity evaluated at the black hole's
horizon \cite{Jacobson93}. It differs from the $A/4$ value by a sum of
intrinsic curvature invariants integrated over a cross section of the
horizon.
For $d=4$ in a generic $(Riemann)^2$ gravity the area law is valid, but not
in a generic $(Riemann)^3$ theory \cite{Visser93}.
We would like however to point out that the above statements are true
in vacuum.
As we shall see  in this paper, when matter couples to gravity,
e.g. when an electromagnetic field is present, the area law is not valid
in a generic four dimensional gravitational theory with $(Riemann)^2$
terms.

We consider the 4-dimensional theory
described by the Lagrangian
\begin{equation}
      {\cal L} = {1\over 16\pi}\sqrt{-g}
    ( R + \alpha R^2 + \beta R_{\mu\nu} R^{\mu\nu} ) + {\cal L}_{M},
\label{QuadraticLagrangian}
\end{equation}
where $R_{\mu\nu}, R$ are the Ricci tensor and the scalar curvature
with respect to the metric  $g_{\mu\nu}$ and $g:= det |g_{ab}|$.
$\alpha$ and $\beta$ are some coupling constants.
We will finally take as the matter part, ${\cal L}_{M}$, the
usual  electromagnetic Lagrangian.

It is known that the gravitational part of such a theory
can be described in terms
of a massless tensorial graviton field, a massive
scalar field if $\alpha\neq 0$, and/or a massive tensor field whenever
$\beta\neq 0$, see \cite{Magnano87}, \cite{Jakubiec88}, \cite{Audretsch93}
and references therein. From the analysis of the field equations at the
linearized level \cite{Audretsch93},\cite{Teyssandier89},
it follows that the ``source'' of the additional
massive fields is, for the scalar massive field, the trace $T^{\rm(M)}$ of
the usual matter stress energy momentum tensor $T^{\rm(M)}_{\mu\nu}$; while
for the massive tensorial field the source is some combination of
 $T^{\rm(M)}_{\mu\nu}$ components.
Using this fact we may understand the motivation and some of the conclusions
of this paper: the electromagnetic field, having a traceless stress energy
momentum tensor cannot excite the scalar massive field of the theory. If,
furthermore, $\beta=0$ then the solution may well not differ from
that in Einstein's theory, namely it may be the RN solution.
Thus the no-hair conjecture seems to hold in this case.
In fact a no-hair theorem can be proven \cite{Whitt84}.
However, it is immediately clear that this cannot be guaranteed if
$\beta\neq 0$.
It is the purpose of this paper to study this last case and to show that
the charged black holes of the theory with Lagrangian given by
(\ref{QuadraticLagrangian}) do not coincide with the RN or, if rotation is
included, with the KN metrics.
We will have to work in some approximation which however in some limit
will allow us to study the thermodynamical properties of these
charged black hole solutions of the quadratic theory
(\ref{QuadraticLagrangian}).
In particular we shall consider only solutions
expandable in small $alpha$ and $beta$.

The paper is organized as follows.
In Section II we present the field equations in
the exact and in some approximate case. In Section III we proceed mainly
with the solution of the approximate field equations in the static and
spherically symmetric case and discuss the
range of the validity of the obtained black hole solution.
This allows us to study in Section IV its thermodynamical properties.
In Sec. V we extend the previous analysis to include rotation.
Finally, in the appendix we give the proof of the basic relations used
for the solution of the field equations.

We use throughout this paper units in which $\hbar=c=G=k_B=1$,
metric signature $(-+++)$, Riemann tensor
$R^a_{bcd}:=-\partial_d\Gamma^a_{bc} + \cdots$
and Ricci tensor $R_{ab}:=R^c_{acb}$. Finally Gaussian
electromagnetic units are employed.

\section{Field Equations}

\subsection{Exact Field Equations}

  The gravitational field equations for the theory in
Eq.\ (\ref{QuadraticLagrangian}) read

\begin{eqnarray}
  && (1+2\alpha R)(R_{\mu\nu} - {1\over 2}g_{\mu\nu}R)
       + {\alpha \over 2}R^2  g_{\mu\nu}                        \cr
  && + (2\alpha +
      {\beta\over 2})g_{\mu\nu} R_{;p}{}^{;p}
      -(2\alpha+\beta) R_{;\mu\nu} + \beta R_{\mu\nu;p}{}^{;p}   \cr
  && - {\beta\over 2} R_{pq}R^{pq}g_{\mu\nu} +
     2\beta R_{pq}R{}_\mu{}^p{}_\nu{}^q                         \cr
  && \quad\quad= {-16\pi\over\sqrt{-g}}
         {\delta  S_M \over \delta (g^{\mu\nu})} :=
  8\pi T_{\mu\nu} ,
\label{FieldEqs}
\end{eqnarray}
where $T_{\mu\nu}$ is the stress-energy-momentum tensor of the matter
fields in the Lagrangian of Eq.\ (\ref{QuadraticLagrangian}).
In this paper we will  consider as matter field the
electromagnetic field $F_{\mu\nu}$, coupled minimally to gravity.
The field equations for $F_{\mu\nu}$ are the Maxwell equations
\begin{equation}
  \sqrt{-g}\nabla_\mu F^{\mu\nu}=
  \partial_{\mu}(\sqrt{-g} F^{\mu\nu} ) = -4\pi\sqrt{-g} j^{\nu} ,
\label{MaxwellEqs}
\end{equation}
where $j^\nu$ is the electromagnetic current.

Because of the complexity of these equations it is extremely difficult
to obtain exact solutions even in cases of high symmetry. Consequently
one should be prepared to work with some approximation scheme.
We shall now discuss the main elements of our approximation method.

\subsection{Approximate Field Equations}

We concentrate on the case where the quadratic theory is slightly
different from Einstein's general relativity. In particular
let us consider the case where we can neglect the quadratic curvature
terms in field equations (\ref{FieldEqs}). These ``linearized''
field equations read
\begin{eqnarray}
G_{ab} &&\approx 8\pi T_{ab}                       \cr
       && - (2\alpha + {\beta\over 2}) g_{ab}\Box R +
       (2\alpha + \beta)R_{;ab} - \beta \Box R_{ab} .
\label{LinearFieldEqs}
\end{eqnarray}
More specifically, this linearization demands that terms
like $R^2, R_{ab}R^{ab}$, which are proportional to $\alpha, \beta$
parameters, can be neglected on the one hand with respect to $R, R_{ab}$
ones and on the other hand with respect to second derivatives of
curvature terms which are also proportional to $\alpha$ and $\beta$
parameters.
The first requirement, can be achieved by choosing sufficiently
small values of $\alpha$ and $\beta$.
Small values for these parameters is actually what nature has
chosen, as it is concluded from all the tests general
relativity ($\alpha=0=\beta$) has so far successfully passed.
The second requirement, however, may not be achievable
everywhere in the spacetime since
it depends on the particular behavior of curvature components.
We will discuss it further below, when we will consider the case
of charged black holes.

   Assuming that the above requirements are satisfied,
the spacetime metric $g_{ab}$ in the quadratic theory
in the presence of some matter field, is slightly
modified from the respective metric $\widehat g_{ab}$
in Einstein's theory with the same matter field. We can write
\begin{equation}
  g_{ab}= \widehat g_{ab} + \chi \widehat g_{ab} + \psi_{ab} ,
\label{QuadDecomposedMetric}
\end{equation}
with $|\chi| \ll 1,\quad |\psi_{ab}| \ll |\widehat g_{ab}|$.
As it is shown in the appendix, by choosing appropriately
$\chi, \psi_{ab}$ the ``linearized'' gravitational field
equations Eqs.\ (\ref{LinearFieldEqs}) are equivalent to
the following system of equations
\begin{eqnarray}
    G_{\mu\nu}(\widehat g_{ab})
          &=& 8\pi T_{\mu\nu} ,
                                            \label{QuadFieldEqI} \\
  (\Box - m_0^2)\chi
          &=& -{8\pi\over 3} T,
                                            \label{QuadFieldEqII} \\
  (\Box- m_1^2)\psi_{\mu\nu}
          &+& \left( 1-{m_1^2\over m_0^2} \right )
            \nabla_{\mu} \nabla_{\nu}
            (\psi^\lambda{}_\lambda{})                            \cr
          &=&
     16\pi\left(T_{\mu\nu} - {1\over 3}T g_{\mu\nu} \right ),
                                             \label{QuadFieldEqIII}
\end{eqnarray}
where
\begin{equation}
  m_0^{-2}=6\alpha+2\beta, \qquad
  m_1^{-2}= -\beta .
\label{MassParameters}
\end{equation}
Finally $\psi_{\mu\nu}$ should satisfy the condition
\begin{equation}
  (\psi_{ab} - \psi^\lambda_\lambda g_{ab})^{;b} = 0 .
\label{PsiCondition}
\end{equation}

In Eqs.\ (\ref{QuadFieldEqI}), (\ref{QuadFieldEqII}) and
(\ref{QuadFieldEqIII}) $G_{\mu\nu}$ is the Einstein tensor
for the metric  $\widehat g_{\mu\nu}$. $T_{\mu\nu}$ is the
stress-energy-momentum tensor  of the matter fields in the original
Lagrangian.
However in our approximation, where
quadratic terms in curvature are neglected,
the energy tensor needs to be constructed using only the
$\widehat g_{ab}$ metric. Terms that are omitted in this way
are $(\partial T_{ab} / \partial g_{\mu\nu} ) \delta g_{\mu\nu} \propto
{\cal O}(curvature^2)$
which although they are linear in $\chi$ and $\psi_{ab}$ fields
they are nevertheless quadratic in curvature.
For the same reason the metric $g_{\mu\nu}$ in the right hand side of
Eq.\ (\ref{QuadFieldEqIII}) can be replaced with $\widehat g_{\mu\nu}$.
Finally, in our appoximation the derivative operators can also be taken
with respect to $\widehat g_{\mu\nu}$ metric.

To have positive mass parameters $m_0, m_1$ we shall hereafter assume
the no tachyon constraints
\begin{equation}
  3\alpha + \beta \ge 0, \quad \beta \le 0 .
\label{NoTachyon}
\end{equation}

The above field equations demonstrate the decomposition of the quadratic
gravitational theory into a theory with a graviton field
$\widehat{g}_{ab}$, a massive scalar field $\chi$
and a massive tensorial field $\psi_{\mu\nu}$,
all of them coupled with the stress energy momentum tensor $T_{\mu\nu}$.
In particular one should notice here what we have remarked in the
introduction: the scalar field $\chi$ has as
source the trace of $T_{\mu\nu}$ and thus will not get excited by the
traceless $T_{\mu\nu}$ of an electromagnetic field. On the other hand,
the massive spin-2 field $\psi_{\mu\nu}$ in general will.
 Finally let us remark that there are some issues
regarding the equivalence of such a decomposition with the initial
theory. These will not be discussed in this paper, see \cite{Audretsch93}
and references therein. Instead we shall only view these equations as a
convenient mathematical reduction step for solving in some approximation
the initial fourth order derivative theory dealing only with second
order derivative equations.

\section{Solutions}
  We are interested in charged black hole solutions of the quadratic
theory with the Lagrangian (\ref{QuadraticLagrangian}).
We will consider both static and stationary solutions.
However, we prefer to discuss in more detail
the static case which will allow  the illustration of our method
and the exposition of our main results without particular
technical complications.
Then, in another section, we will extend the analysis to include
rotation.

We consider now the case where the matter field is static and
spherically symmetric generated by an electric charge $Q$ centered
at the origin $r=0$.
The respective spacetime will be spherically symmetric and static and
therefore there is a coordinate system where the metric can be written as
\begin{equation}
    ds^2 = -B(r)\,dt^2 + A(r)\,dr^2 + r^2\left ( d\theta^2 +
         \sin^2\theta d\phi^2 \right ) ,
\label{SpherSymMetric}
\end{equation}
Here $\theta$ and $\phi$ are the usual spherical angular coordinates.

The field equations for the Maxwell field $F_{\mu\nu}$ are given by
Eq.\ (\ref{MaxwellEqs}) where
the current source has only a time component equal to
$j^0 = Q\delta(r)/( 4\pi r^2 \sqrt{AB})$.
These electromagnetic field equations  are easily solved in the
background metric (\ref{SpherSymMetric}). It turns out that the only
non vanishing components of the $F^{\mu\nu}$ tensor are the $tr$ ($rt$)
ones with
\begin{equation}
  F^{tr}= -{Q\over r^2 \sqrt{AB} } .
\label{MaxwellSolution}
\end{equation}

The stress energy momentum tensor for this field configuration is
\begin{equation}
  T^t_t=T^r_r=-T^\theta_\theta=-T^\phi_\phi =-{Q^2\over 8\pi r^4} .
\label{SEMMaxwell}
\end{equation}

We now look for solutions to the gravitational equations.

\subsection{Exact Static Black Hole Solutions}

{\it Case} (i): $\alpha=0, \beta=0$.
In this case we have Einstein's theory where the solution is known to
be unique.
It is the Reissner-Nordstr\"om (RN) black hole metric given by
Eq.\ (\ref{SpherSymMetric}) with
\begin{equation}
  B(r)=A(r)^{-1}= 1- {2 M\over r} + {Q^2\over r^2} .
\label{RNmetric}
\end{equation}
where $M$ and $Q$ are respectively the mass and the total charge
of the black hole.

{\it Case} (ii): $\alpha\neq 0, \beta=0$
Here we can check directly \cite{Mathematica}
that the general relativistic RN black hole
metric is also an exact solution for this theory. The question is whether
it is unique. The answer seems to be yes
since according to Whitt \cite{Whitt84}
a no-hair theorem can be proven for
theories with gravitational Lagrangian $L=R + \alpha R^2$ in vacuum
and in the presence of electromagnetic matter in the case
where the condition $R\neq -\frac{1}{2\alpha}$ holds everywhere.

{\it Case} (iii): $\beta\neq 0$.
Now the direct check with the RN solution \cite{Mathematica},
shows that the RN metric is not
a solution, except for the vacuum case with zero charge $Q=0$ where one
solution is known to be the general relativistic Schwarzschild black hole.
The field equations are very difficult and have not allowed us to
find an exact solution.

\subsection{Approximate Static Black Hole Solutions}

We assume now that there exists some parameter space where one may use
the approximate system of second order field equations
(\ref{QuadFieldEqI})-(\ref{QuadFieldEqIII}) instead
of the original fourth order derivative equations. After finding
solutions to these equations one has to check whether they are
compatible with the assumptions that we have made when deriving
these field equations.

For the stress energy momentum tensor of Eq.\ (\ref{SEMMaxwell})
the solution for  $\widehat{g}_{\mu\nu}$ in Eq.\ (\ref{QuadFieldEqI})
is of course the RN metric given by Eqs.\ (\ref{SpherSymMetric}) and
(\ref{RNmetric}).
Now, in the background of this $\widehat{g}_{\mu\nu}$ metric we have to
solve Eqs.\ (\ref{QuadFieldEqII}) and (\ref{QuadFieldEqIII}).
Notice, once again, that the $\chi$ field is not excited because $T=0$.
Taking into account the symmetries of our problem we are left
with the following set of equations for the field $\psi_{\mu\nu}$.
\begin{equation}
  (\widehat{\nabla}^2- m_1^2)\psi_{\mu\nu} \approx 16\pi T_{\mu\nu},
\label{psiEqs}
\end{equation}
where $\widehat{\nabla}^2$ is with respect to the RN metric,
and  the stress energy momentum tensor $T_{\mu\nu}$ is given by
Eqs.\ (\ref{SEMMaxwell}).

To proceed  we shall consider the
case of sufficiently small $|\beta|$ parameter
such that $M\gg m_1^{-1}$ (remember $m_1^{-2}:=-\beta$).
This implies that with $r\geq M$ we will also have $r\gg m_1^{-1}$.
Using this fact we observe that we can obtain the first leading
term for $\psi_{\mu\nu}$ very simply.
Indeed, in the left hand side of the field equations (\ref{psiEqs})
the mass term dominates over the derivative terms for
$r\geq M\gg m_1^{-1}$.
Thus, to a good approximation we can neglect the derivative terms and
arrive at the following diagonal solution for $\psi_{\mu\nu}$
\begin{eqnarray}
  \psi_{tt} &\approx& - {2Q^2\over m_1^2 r^4}
      (1-{2M\over r} + {Q^2\over r^2}), \cr
  \psi_{rr} &\approx&  {2Q^2\over m_1^2 r^4}
      (1-{2M\over r} + {Q^2\over r^2})^{-1}, \cr
  \psi_{\theta\theta} &\approx&
       -{2Q^2\over m_1^2 r^2}, \cr
  \psi_{\phi\phi} &\approx&
       -{2Q^2\over m_1^2 r^2}\sin^2\theta .
\label{psiSolution}
\end{eqnarray}
With a straightforward calculation one can check that the condition
(\ref{PsiCondition}) is satisfied.

Finally replacing in Eq.\ (\ref{QuadDecomposedMetric}) the components
of the Reissner-Nordstr\"om metric $\widehat{g}_{\mu\nu}$ and of the
$\psi_{\mu\nu}$ field from the last equation (\ref{psiSolution}) we
obtain the metric $g_{\mu\nu}$ for the initial fourth order
gravitational theory. This can be written as
\begin{eqnarray}
  ds^2 \approx &&
   - \left( 1-{2M\over r}  + {Q^2\over r^2} \right)
     \left( 1+ {2Q^2\over m_1^2 r^4} \right) dt^2                 \cr
    && + \left( 1-{2M\over r}  + {Q^2\over r^2} \right)^{-1}
         \left( 1+ {2Q^2\over m_1^2 r^4} \right) dr^2             \cr
    && + r^2 \left( 1- {2Q^2\over m_1^2 r^4} \right)
         (d\theta^2 + \sin^2\theta d\phi^2) .
\label{ApproxSol}
\end{eqnarray}

It is easy to see that in this metric the ``Newtonian'' gravitational
potential is modified. At large radial distances, particles will feel an
additional repulsive force  $4Q^2/(m_1^2 r^5)$ besides the force that
they feel in the RN spacetime. Note however that null geodesics on the
$\theta=0$ plane coincide with those of the RN metric.

Up to now we have solved the approximate field equations
(\ref{QuadFieldEqI})-(\ref{QuadFieldEqIII})
in the case where $r\geq M\gg m_1^{-1}$.
However, we still have to check whether the conditions for using
the approximate field equations are satisfied.
With straightforward but tedious calculations, we find that
these conditions are satisfied for the solution (\ref{ApproxSol})
at sufficiently large distances from the black hole's horizon $r\gg M$.
However, if one is interested in a solution that is approximately valid
in the region from radial infinity down to the horizon, one has still
to restrict it to sufficiently small charges $|Q| \ll M$.
We will provide here only a dimensional argument which leads to
this condition. Remember that for using the approximate field equations
we needed the curvature squared terms to be negligible with respect to
second derivatives of curvature terms. For our particular problem
the metric is approximately the Reissner-Nordstr\"om one, and thus
a typical curvature term will be $\propto Q^2/r^4$. Therefore we must
have $(Q^2/r^4)^2 \ll (1/r^2)(Q^2/r^4)$. If this condition is to be
satisfied also near the horizon $r\approx M$ one gets the restriction
$Q^2\ll M^2$.
And this is what we assume in the following section where we discuss
the thermodynamic properties of the approximate solution
(\ref{ApproxSol}).

\section{Thermodynamic Properties of the Solution}

As we have seen in the previous section the metric of a charged black
hole in the quadratic gravitational theory (\ref{QuadraticLagrangian})
is not the Reissner-Nordstr\"om solution. To a good approximation
the metric is given by Eq.\ (\ref{ApproxSol}). Its range of validity
extends down to the vicinity of the horizon  provided that
$M\gg |Q|$ and $M \gg m_1^{-1}$. This will allow us to calculate
some of the thermodynamic quantities associated to black holes.
In particular, we are interested in the entropy-area relationship.
Since the metric is known to be spherically symmetric and static,
it is easy to employ standard Euclidean techniques to obtain the
temperature associated to the black hole and an expression for its
entropy \cite{Gibbons75}.
However the same results are obtained with a more simple minded
thermodynamical treatment \cite{Bekenstein73}.
We prefer here to follow this last method.

With the argument due to Hawking, which is based on a
semiclassical calculation of quantum effects near the horizon,
one can associate to a black hole a temperature $T=\kappa/(2\pi)$
where $\kappa$ is the surface gravity of the black hole's horizon.
Note that for our metric the horizon is at a radial coordinate
$r_{\rm H}$ given by
\begin{equation}
  r_{\rm H} =  M  + \sqrt{ M^2 - Q^2 } .
\label{rHorizon}
\end{equation}
The surface gravity of the horizon $\kappa(r_{\rm H})$ is then
easily computed
\begin{equation}
  \kappa(r_{\rm H}) =
  -{1\over 2} {\partial g_{tt}/\partial r \over \sqrt{ -g_{tt} g_{rr} } }
  = { \sqrt{M^2 - Q ^2} \over (M + \sqrt{M^2 - Q ^2})^2 } .
\label{SurfaceGravity}
\end{equation}
We notice here that $\kappa(r_{\rm H})$ has the same value as for the
RN black hole of general relativity. Consequently the same is true for the
temperature.

This coincidence should not create the wrong impression that everything
regarding the thermodynamics-geometry relationship for our black hole
is the same as for the General Relativity.
On the one hand the coincidence itself is probably accidental in our
approximation and does not survive in the exact solution. On the other
hand, even to our approximation, the other important geometric quantity
of black hole, namely its area $A$, has not the same expression in
the RN solution and in our solution given by Eq.\ (\ref{ApproxSol}).
In fact the horizon area of our quadratic gravity black hole is given by
\begin{equation}
  A = 4\pi r_{\rm H}^2 \left ( 1 - {2Q^2 \over m^2 r_{\rm H}^4} \right ) .
\label{Area}
\end{equation}

By replacing into this expression the $r_{\rm H}$ of Eq.\ (\ref{rHorizon})
and solving the resulting expression for the mass $M$ of the black hole we
obtain the fundamental relation of black hole thermodynamics:
\begin{equation}
  M= {
       \left [ {A \over 8\pi} + \sqrt{\left( {A \over 8\pi}\right )^2 +
            {2Q^2 \over m^2}} + Q^2 \right ]
     \over
       2 \left [ {A \over 8\pi} + \sqrt{\left({A \over 8\pi}\right)^2 +
            {2Q^2 \over m^2}} \right ]^{1/2}
     } .
\label{FundRel}
\end{equation}

By differentiation of this equation we obtain the first law of black
hole thermodynamics
\begin{equation}
    dM = {\partial M \over \partial A}\Bigl|_Q dA +
       {\partial M \over \partial Q}\Bigl|_A dQ
       := T_{\rm H} dS + \Phi_{\rm H} dQ ,
\label{FirstLaw}
\end{equation}
where $S$ and $\Phi$ represent, respectively, the entropy and the electric
potential at the horizon of the black hole.

Now the task is the proper identification of the quantities
$T_{\rm H}, S, \Phi_{\rm H}$ and $Q$ in (\ref{FirstLaw}) with the black
hole parameters.
The derivatives appearing in (\ref{FirstLaw}) can be explicitly obtained
using Eq.\ (\ref{FundRel}).
The temperature $T_{\rm H}$ is naturally identified with the
Hawking temperature $T=\kappa(r_{\rm H})/(2\pi)$ that we have calculated
above.
Similarly $Q$ is the total electric charge of the black hole, i.e. the
same quantity that appears in our quadratic black hole metric.
On the other hand, the electromagnetic potential $\Phi_{\rm H}$
on the horizon can be computed by direct use of Maxwell equations
\begin{equation}
    \Phi_H = \int_{r_{\rm H}}^\infty {Q\over r^2}
             (1 - {Q^2\over m_1^2 r^4})dr
             ={Q \over r_{\rm H}} + {\cal O}({Q^3 \over m_1^2 r^5}) .
\label{ElPot}
\end{equation}
Thus by inserting  Eqs. (\ref{SurfaceGravity}) and
(\ref{ElPot}) into Eq.\ (\ref{FirstLaw})
we can, upon integration, identify the black hole entropy $S$ as
\begin{equation}
  S \approx {A\over 4} + {8\pi^2 Q^2 \over m_1^2 A}
    \approx \pi r_{\rm H}^2 .
\label{Entropy}
\end{equation}

Here we observe that extra terms appear when we consider $S$ as a
function of $A$. Thus in general, the simple relationship
$S={1\over 4} A$ no longer holds in quadratic gravitational
Lagrangian theories.
Notice however that in terms of the black hole parameters $M$ and $Q$,
the entropy $S$ has the same value as in RN black hole. The same is
true, as we have seen above, for $T_{\rm H}$ and $\Phi_{\rm H}$ and
also for other derived quantities as e.g. the heat capacity since
\begin{equation}
  C_{Q} = T {\partial S \over \partial T}\Bigl \vert _Q =C_Q^{\rm RN}.
\label{HeatCapacity}
\end{equation}
Let us remark once again that this nice coincidence probably
happens only in our approximation and will not survive in
the exact quadratic charged black hole solution.

\section{Extension to the rotating case}

In this section we will extend our calculations to the case of rotating
charged black holes in the quadratic theory (\ref{QuadraticLagrangian}).
We will work in the approximation where we can use the
field equations (\ref{QuadFieldEqI})-(\ref{QuadFieldEqIII}).
We will use the same methods that we used in the previous
sections for the static case.

The metric can now be written in the form

\begin{equation}
    ds^2 = g_{tt} dt^2 + 2g_{t\phi} dt d\phi + g_{rr} dr^2 +
         g_{\phi\phi}d\phi^2 + g_{\theta\theta} d\theta^2 .
\label{RotTemplate}
\end{equation}

As in the static case we can almost immediately solve the approximate
field equations. Indeed, $\widehat{g}_{\mu\nu}$ will be the known general
relativistic Kerr-Newman rotating black hole metric.
The scalar field $\chi$ will not get excited since the source is the
traceless electromagnetic stress energy momentum tensor.
Therefore, we are left with the field equations for the massive scalar
field $\psi_{\mu\nu}$ which are to be solved in the background of the KN
metric.
Then with the same arguments as in static case we find that the solution
for $\psi_{\mu\nu}$ is approximately given by
\begin{equation}
  \psi_{\mu\nu}\approx -{16\pi \over m_1^2} T_{\mu\nu}^{\rm (KN)},
\label{psiRotEq}
\end{equation}
where $T_{\mu\nu}^{\rm (KN)}$ is the energy tensor of the
Kerr-Newman black hole.
Finally the metric in the quadratic theory is given by
\begin{equation}
  g_{\mu\nu} \approx \widehat{g}_{\mu\nu} + \psi_{\mu\nu}.
\label{RotMetric}
\end{equation}

Using the metric tensor and the $T_{\mu\nu}^{\rm (KN)}$ tensor of the
Kerr-Newman black hole in Boyer-Lindquist coordinates \cite{Gravitation},
we obtain the metric $g_{\mu\nu}$

\begin{eqnarray}
    g_{tt} &=& -{\Delta I_{(+)} \over \rho^2} +
        {a^2 \sin^2\theta \over \rho^2 } I_{(-)},          \cr
    g_{t\phi} &=& [\Delta I_{(+)} - (\rho^2+a^2)I_{(-)}]
        {a \sin^2\theta \over \rho^2 },                     \cr
    g_{\phi\phi} &=& [(r^2+a^2)^2I_{(-)}-
             \Delta a^2 \sin^2\theta I_{(+)}]
             {a \sin^2\theta \over \rho^2 },                \cr
    g_{\theta\theta} &=& \rho^2 I_{(-)},                    \cr
    g_{rr} &=& {\rho^2 \over \Delta}I_{(+)},
\label{RotQuadMetric}
\end{eqnarray}
where
\begin{equation}
  I_{(\pm)} = (1 \pm {2Q^2 \over m_1^2 \rho^4} ),
\end{equation}
and $\rho:= r^2 + a^2 cos^2\theta, \Delta:= r^2 - 2Mr + a^2 + Q^2$.
Here $a$ is the usual angular momentum per total mass parameter.

As in the static case we shall take $\vert Q\vert\ll M$ to be able to
discuss the thermodynamic behavior of the solution.
Having the metric it is now straightforward to calculate the radius of
the event horizon $r_{\rm H}$, the surface gravity $\kappa(r_{\rm H})$
and the horizon's angular velocity $\Omega_{\rm H}$.
All of them turn out to have the general relativity values namely
\begin{equation}
  r_{\rm H}= M + \sqrt{M^2- Q^2 - a^2},
\label{RotHorizonRadius}
\end{equation}

\begin{equation}
  \kappa(r_{\rm H}) = { \sqrt{M^2-a^2-Q^2} \over r_{\rm H}^2 + a^2 },
\label{RotSurfaceGravity}
\end{equation}

\begin{equation}
  \Omega_{\rm H} = {a \over r_{\rm H}^2 + a^2}.
\label{HorAngulVel}
\end{equation}
Straightforward is also the calculation for the area $A$ of the event
horizon

\begin{equation}
    A = 4\pi \left\lbrace (r_{\rm H}^2 + a^2 ) +
        {Q^2 \over m_1^2 r_{\rm H}^2}
        \left[ 1 + {(r_{\rm H}^2 + a^2)\over a r_{\rm H}} {\rm arctg}
        ({a\over r_{\rm H}}) \right]
        \right\rbrace .
\label{RotHorizonArea}
\end{equation}
Note that, as in the static case, this is different from the general
relativity result. From this equation we can obtain $M$
as a function of $A,$ $Q$ and $J$.

The first law of black hole thermodynamics now reads
\begin{eqnarray}
    dM =&& {\partial M \over \partial A}\Bigl|_{Q,J} dA +
       {\partial M \over \partial Q}\Bigl|_{A,J} dQ +
       {\partial M \over \partial J}\Bigl|_{A,Q} dJ                  \cr
       && := T_{\rm H} dS + \Phi_{\rm H} dQ + \Omega_{\rm H} dJ ,
\label{RotFirstLaw}
\end{eqnarray}
where the new quantities not present in the static case are the horizon's
angular velocity $\Omega_{\rm H}$ and black hole's angular momentum $J$.
The partial derivatives in (\ref{RotFirstLaw}) can be explicitly
calculated, and the $T_{\rm H}, Q, \Phi_{\rm H}$ are naturally
identified as in the static case.
Hence we are left with a differential expression for the calculation of
the entropy $S$ in terms of the parameters $A, J$, and $Q$.
Upon integration of this expression we find
\begin{equation}
     S(A,J,Q) = {A\over 4} + {8\pi^2 Q^2 \over m_1^2 A} +
       {8 (4\pi)^4 Q^2 J^2 \over 3 m_1^2 A^3} + ...
\label{RotEntropy}
\end{equation}

We observe that in general the entropy will not be a function only of the
area of black hole, but also of the other parameters that characterize it.

It is interesting to notice that also in the rotational case,
the actual value of the entropy, that is the one in terms of $M,Q,J$,
is the same as in general relativity since $S=4\pi r_{\rm H}^2$ still holds
in our approximation.
But probably this is not true in the exact solution.

\section{Summary and Discussion}

We have looked for solutions representing electrically
charged, static and stationary black holes in the theory with the
quadratic gravitational Lagrangian of Eq.\ (\ref{QuadraticLagrangian}).
The gravitational field equations which involved fourth order
derivatives for the metric have been reduced to a set of field equations
with second order derivatives for a massless tensorial field
together with a massive scalar and a massive tensorial field.
Solving approximately these equations we have seen that
the modifications from the general relativistic results come only from
the massive tensorial field. The scalar field does not contribute since
its source is the trace
of the matter stress energy momentum tensor which in our case is the
traceless electromagnetic energy tensor. This is in fact an exact result
since, as we have directly checked, the Reissner-Nordstr\"om solution is
an exact solution of the initial quadratic gravitational field equations
only if the massive tensorial field is missing ($\alpha\neq 0, \beta=0$
in Eq.\ (\ref{QuadraticLagrangian})).

Computing several thermodynamic quantities for our black
hole solutions we have found that
these quantities retain, in our approximation, their general relativistic
value.
This coincidence will quite probably not survive in the exact solution.
However it is interesting to note here that for generalized charged black holes
in other theories usually one finds corrections in the temperature and other
thermodynamic quantities, see Ref. \cite{Mignemi93} for a discussion in the
context of string theory and Ref. \cite{Tomimatsu88} in terms of Kaluza-Klein
theories.

Finally, for our black hole solutions we have found that their entropy
does not follow the simple area law $S= A/4$. Such a behavior is known
for theories  that contain non linear curvature terms in their gravitational
Lagrangian and have dimensionality higher than four. The respective
investigations
have been mainly performed with the Lovelock Lagrangian \cite{Jacobson93}.
To have in this theory non trivial results in four dimensions one should have
an
effective non trivial coupling of the curvature to some other field,
see e.g. Ref. \cite{Tomimatsu88} in the context of Kaluza-Klein theories.
As we have shown in this paper, we can have in four dimensions
a violation of the area law just by including a minimally coupled
electromagnetic
field in the quadratic gravitational Lagrangian of Eq.\
(\ref{QuadraticLagrangian}).

\section{Appendix}

The field equations of the quadratic theory
in the approximation where only linear terms in curvature tensors are
retained, are given by equation (\ref{LinearFieldEqs}).

Let us now consider the following metric transformation
\begin{equation}
  g_{ab} \to \widehat g_{ab} = g_{ab} - \chi g_{ab} - \psi_{ab},
\label{MetricTransformation}
\end{equation}
with $\chi$ and $\psi_{ab}$ as scalar and tensorial field perturbation,
that is with $\vert\chi\vert\ll 1$ and $\psi_{ab} \ll g_{ab}$.
Hereafter, hatted quantities will refer to the $\widehat{g}_{ab}$ metric.

For the purposes of the paper we will keep only linear terms in $\chi$
and $\psi_{ab}$ fields. Then  the transformation in
Eq.\ (\ref{MetricTransformation}) implies that the Einstein tensor
$\widehat G_{ab}$ is related to $G_{ab}$ with
\begin{eqnarray}
    \widehat{G}_{ab} = G_{ab}
    &+& {1\over 2} \Box \psi_{ab} +
     {1\over 2} (\psi{}_p{}^p{}_{;ab} - \psi{}_{pa;b}{}^{;p} -
      \psi{}_{pb;a}{}^{;p})                                      \cr
    &+& {1\over 2} g_{ab} ( \psi{}_{pq}{}^{;pq}
      - \Box \psi{}_p{}^p)                                    \cr
    &+& \chi{}_{;ab} - (\Box \chi ) g_{ab}
      + O(\psi^2, \chi^2) .
\label{EinsteinTransf}
\end{eqnarray}
Here the derivative operators, being unhatted, are with respect to
the $g_{ab}$ metric.

Substituting Eq.\ (\ref{LinearFieldEqs}) in Eq.\ (\ref{EinsteinTransf})
we obtain (to linear order)
\begin{eqnarray}
  \widehat{G}_{ab}
      &\approx& 8\pi T_{ab}                             \cr
      && - (2\alpha + {\beta\over 2})
               g_{ab}\Box R + (2\alpha + \beta)R_{;ab} -
	       \beta \Box R_{ab}                                 \cr
      && + {1\over 2}\Box \psi_{ab} + {1\over 2}
	      (\psi{}_p{}^p{}_{;ab} -
               \psi{}_{pa;b}{}^{;p} - \psi{}_{pb;a}{}^{;p})     \cr
	  && +  {1\over 2} g_{ab} ( \psi{}_{pq}{}^{;pq} -
	      \Box \psi{}_p{}^p)                                 \cr
      && + \chi{}_{;ab} - (\Box \chi ) g_{ab}
		+ O(\psi^2, \chi^2).
\label{LinFieldEqs}
\end{eqnarray}

Our aim now is to decrease the order of metric derivatives that appear
in the field equations (\ref{LinFieldEqs}).
This can be  done by  selecting  the tensorial field $\psi_{ab}$ and the
scalar field $\chi$ in such a way that the terms on the r.h.s. of
Eq.\ (\ref{LinFieldEqs}) which contain higher order than second metric
derivatives cancel out.
Obviously, with $\psi_{ab}$ we can cancel the term $\Box R_{ab}$.
For this let us choose
\begin{equation}
  \psi_{ab} = 2\beta R_{ab} + \lambda R g_{ab},
\label{psiChoice}
\end{equation}
where $\lambda$ is some constant.
Then it follows

\begin{eqnarray}
   \widehat{G}_{ab} = 8\pi T_{ab}
    &-& (2\alpha+\beta+\lambda) g_{ab} \Box R
        + (2\alpha + \beta + \lambda) R_{;ab}                   \cr
    &+& \chi_{;ab} - g_{ab}\Box\chi .
\label{simplifiedFieldEqs}
\end{eqnarray}
In deriving Eq.\ (\ref{simplifiedFieldEqs}) we have made use of the
Bianchi identity $R_{ab}{}^{;b}={1\over 2} R_{;a}$ and while commuting
covariant derivatives we have neglected quadratic terms in the
curvature.

Now with the obvious choice
\begin{equation}
  \chi = - (2\alpha + \beta + \lambda) R,
\label{chiChoice}
\end{equation}
we obtain from Eq.\ (\ref{simplifiedFieldEqs}) that
\begin{equation}
  \widehat{G}_{ab} \approx 8\pi T_{ab},
\label{gravitonPart}
\end{equation}
which is indeed an equation with only second derivatives in the metric
$\widehat{g}_{ab}$.
Note that the constant $\lambda$ remains unspecified and can be chosen
arbitrarily.
The field equations for $\chi$ can be derived from the trace of
Eq.\ (\ref{LinearFieldEqs})
\begin{equation}
  (6\alpha + 2\beta)\Box R - R = 8\pi T .
\label{chiFieldEqA}
\end{equation}
With the choice of $\lambda$
\begin{equation}
  \lambda = -{\beta\over 3},
\label{lambdaChoice}
\end{equation}
and using Eq.\ (\ref{chiChoice}) we can write Eq.\ (\ref{chiFieldEqA}) as
\begin{equation}
  \Box\chi - m_0^2\chi = -{8\pi\over 3} T,
    \qquad    m_0^{-2}=6\alpha + 2\beta .
\label{chiFieldEq}
\end{equation}

The field $\psi_{ab}$ should satisfy a set of equations resulting
from the Bianchi identity
\begin{equation}
  (\psi_{ab} - {\beta + \lambda \over 2(\beta + 2\lambda)}
  \psi_p{}^p g_{ab} )^{;b} = 0 .
\label{psiNewConstraint}
\end{equation}

The field equations for $\psi_{ab}$ can be obtained from
Eq.\ (\ref{LinearFieldEqs}) and Eq.\ (\ref{psiChoice})
\begin{eqnarray}
    (\Box - m_1^2)\psi_{ab} =
     && 16\pi T_{ab}
        + (\lambda - 4\alpha - \beta)\Box R g_{ab}           \cr
     && + ({\lambda\over \beta} + 1) R g_{ab}
        + 2(2\alpha +\beta) R_{;ab} .
\label{psiPreEqs}
\end{eqnarray}

With the choice (\ref{lambdaChoice}) and
using Eqs.\ (\ref{chiChoice}), (\ref{chiFieldEq}) and (\ref{psiChoice})
we can write Eq.\ (\ref{psiPreEqs}) as
\begin{eqnarray}
   (\Box - m_1^2)\psi_{ab} - (1 - {m_1^2\over m_0^2}\psi_p{}^p)_{;ab}
   && \cr
    = 16\pi (T_{ab} - {1\over 3} T g_{ab}) . &&
\label{psiFieldEqs}
\end{eqnarray}

   Thus our initial field Eqs.\ (\ref{LinearFieldEqs}) are equivalent
to the system of Eqs.\ (\ref{gravitonPart}), (\ref{chiFieldEq}),
and (\ref{psiFieldEqs}).
In these equations the  $\Box$ operator is with respect to the
metric $g_{ab}$ but it can be also taken with respect to
$\widehat g_{ab}$  in our approximation where we keep only linear terms
in $\chi, \psi_{ab}$, (and equivalently, because of
Eqs.\ (\ref{psiChoice}) and (\ref{chiChoice}),
linear terms in $R, R_{ab}$).  This is the system of field equations
used in the body of the paper.

Note here that Eq.\ (\ref{chiFieldEq}) is not independent since it is
just the trace of Eq. (\ref{psiFieldEqs}). From the physical point of
view, however, it is interesting since it clearly shows that,
at least in our approximations,
the massive scalar sector of the quadratic theory is missing whenever
the trace of the stress energy momentum tensor of matter fields $T$
is zero.

The case of weak gravitational limit is contained in our results
and one may compare with the results of Teyssandier\cite{Teyssandier89}
which however are only valid in a conveniently chosen coordinate system.

\acknowledgments
 This work was supported by the European Community DG XII
 and the Alexander von Humboldt foundation.

\vfil\eject

\end{document}